\documentclass{article}

\usepackage{microtype}
\usepackage{graphicx}
\usepackage{subfigure}
\usepackage{booktabs} 
\newcommand{\xhdr}[1]{\noindent{{\bf #1.}}}
\usepackage{xspace}
\usepackage{mathtools}  
\usepackage[ruled,vlined]{algorithm2e}
\usepackage{amsfonts}
\usepackage{amsmath}
\usepackage{amssymb}
\usepackage{adjustbox}
\newcommand{\name}{\textsc{scGNN}\xspace}

\usepackage{hyperref}

\usepackage[accepted]{icml2019}

\icmltitlerunning{\name}

\begin{document}

\twocolumn[
\icmltitle{\name: scRNA-seq Dropout Imputation \\ via Induced Hierarchical Cell Similarity Graph}

\icmlsetsymbol{equal}{*}

\begin{icmlauthorlist}
\icmlauthor{Kexin Huang}{to}
\end{icmlauthorlist}

\icmlaffiliation{to}{Health Data Science, Harvard T.H. Chan School of Public Health, MA 02120}

\icmlcorrespondingauthor{Kexin Huang}{kexinhuang@hsph.harvard.edu}

\icmlkeywords{Machine Learning, ICML}

\vskip 0.3in
]

\printAffiliationsAndNotice{}

\begin{abstract}
Single-cell RNA sequencing provides tremendous insights to understand biological systems. However, the noise from dropout can corrupt the downstream biological analysis. Hence, it is desirable to impute the dropouts accurately. In this work, we propose a simple and powerful dropout imputation method (\name) by applying a bottlenecked Graph Convolutional Neural Network on an induced hierarchical cell similarity graph. We show \name has competitive performance against state-of-the-art baselines across three datasets and can improve downstream analysis. 
\vspace{-4mm}
\end{abstract}

\section{Introduction}
Single-cell RNA sequencing (scRNA-seq) revolutionizes the ability to study the biological processes by narrowing the study resolution to single cell level~\cite{usoskin2015unbiased,keren2017unique,stephenson2018single, gladka2018single}. 

Despite tremendous success, scRNA-seq suffers from noisiness due to the low RNA capture rate~\cite{kharchenko2014bayesian,jia2017accounting}. Lots of missing values (zero) are observed in scRNA experiment and these values could be either truly zero RNA expression or false negatives due to the low capture rate and failure of amplification. The false negatives is called dropout events. These dropout could potentially affect the downstream analysis such as cell clustering~\cite{hicks2018missing}. Hence, to leverage the scRNA-seq for biological tasks, it is of tremendous necessity and importance to impute these dropouts accurately. 

Several scRNA-seq imputation methods have shown initial success~\cite{arisdakessian2019deepimpute,van2018recovering,huang2018saver,gong2018drimpute} and there are two principles that are leveraged. The first one is the gene co-expression pattern. Through gene regulations, several genes can be expressed together. Hence, if a gene is an incorrectly dropout event, it could be corrected after observing high values of co-expressed genes in the same cell. The second one is cell-cell similarity. As scRNA experiments sequence large number of cells where genes in the similar cells are expected to have similar expression values, a dropout event in one cell can be recovered by imputing the values from a similar cell. These two principles govern the majority of the methods such as MAGIC~\cite{van2018recovering}, SAVER~\cite{huang2018saver}, DCA~\cite{eraslan2019single} and etc. The auto-encoder approaches are reported to have competitive performances in several recent studies~\cite{arisdakessian2019deepimpute,eraslan2019single}. 

Despite promising results, there are still several challenges that could be tackled in previous works:

\textbf{1. Hierarchical cell similarity is missing.} For cell similarity, previous works usually select a set of similar cells and then treat all of them as the same. However, it is likely that there is a hierarchy among the similar cells, i.e. a set of cells have a higher similarity or importance to impute than others. 

\textbf{2. Method that combine cell similarity and gene co-expression principles is lacking.} We observed that most of the previous works focus on improving either cell similarity or gene similarity principle. 

To tackle the above challenges, in this work, we propose a new method called \name, which successfully addresses the above two challenges via:

\textbf{1. An induced hierarchical cell similarity graph.} The induced cell similarity graph is constructed in a way that preserves the cell similarity hierarchy such that the immediate neighbor of a cell is the top K most similar cells, with the n-hop neighbors' similarity decreasing as n goes larger. This network is induced from the raw scRNA-seq data and does not require additional sources.   

\textbf{2. A novel bottle-necked GCN network to integrate cell similarity and gene co-expression patterns.} \name leverages the neighborhood message passing mechanism of GCN. To allow GCN to capture the gene co-expreesion pattern, we inject the bottleneck effect into the GCN network, which has shown efficacy to capture gene co-expressions~\cite{arisdakessian2019deepimpute}. \name achieve this through a simple modifications by manipulating the weight matrix of GCNs. A one line equation is derived to conduct the entire process from raw data to imputed data. 

\name has competitive performances in dropout imputation on three real world datasets when compared to several state-of-the-art baselines. The imputed scRNA matrix reduces noise and improves downstream biological analysis.  

\begin{figure*}[t]
    \centering
    \includegraphics[width=0.65\textwidth]{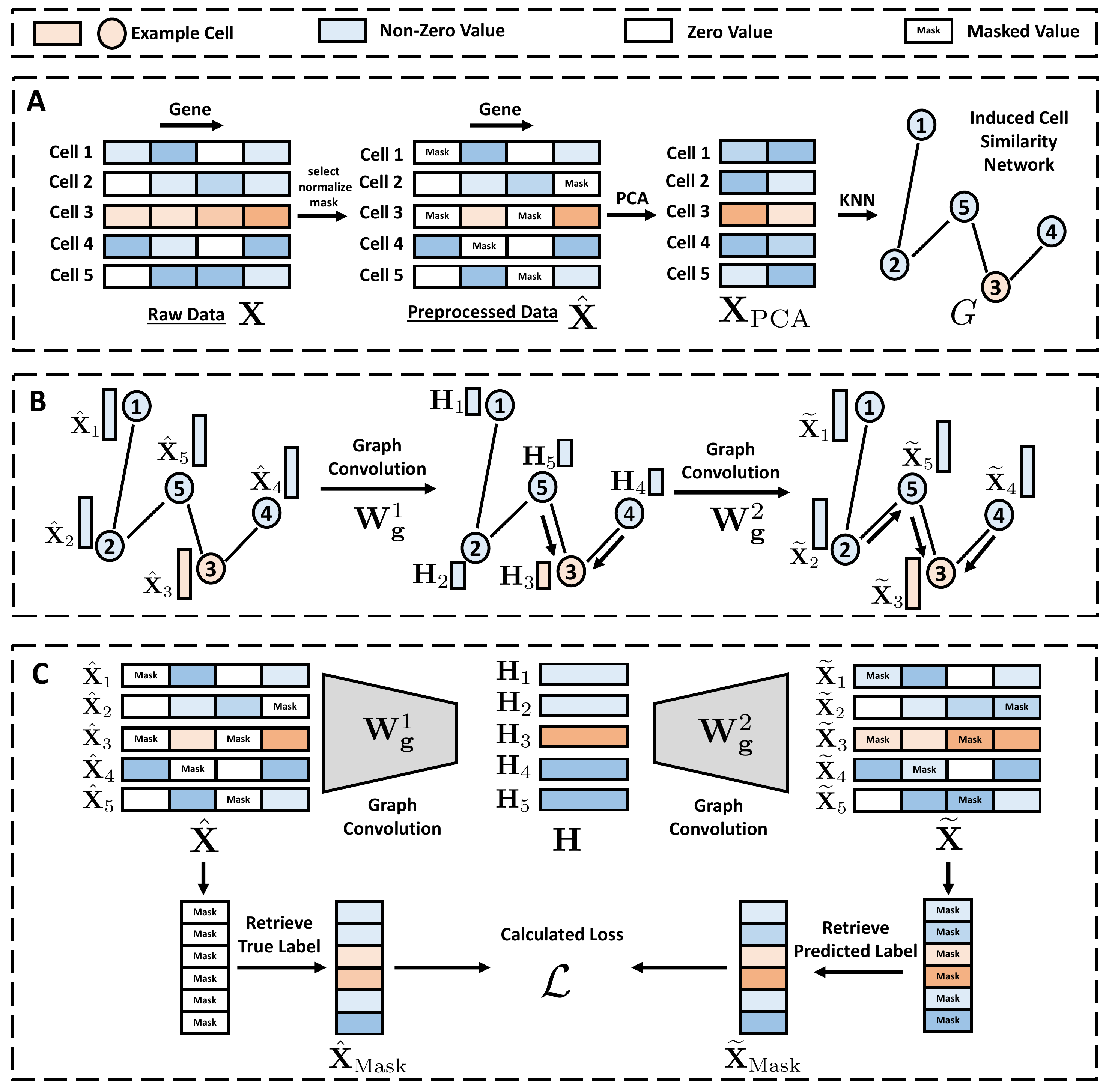}
    \vspace{-2mm}
    \caption{Model Illustration. \textbf{A}. The raw scRNA-seq data is first preprocessed and then we construct the cell-similarity graph $G$ by applying KNN on the PCA embeddings. \textbf{B}. A Graph Convolutional Network is applied on top of $G$. In the first layer, imputation information from the most similar cells (4, 5) are propagated to the center cell (3). In the second layer, the second-order similar cell (2) imputation information is also propagated to the center cell (3) with a discount factor. This models the hierarchical cell similarity imputation principle. Simultaneous to the propagation, in the first layer of GCN, the input scRNA seq data is mapped to low-dimensional latent representation through $\mathbf{W}_\mathbf{g}^1$ (e.g. $\hat{\mathbf{X}}_3 \rightarrow \mathbf{H}_3$). Then, in the second layer of GCN, it is mapped back to the original scRNA dimension through $\mathbf{W}_\mathbf{g}^2$ (e.g. $\hat{\mathbf{X}}_3 \rightarrow \mathbf{H}_3$). This process enables \name to recognize gene co-expression patterns and recover it. \textbf{C}. After imputation, we retrieve the imputed masked values $\widetilde{\mathbf{X}}_\mathrm{Mask}$ and the input masked values $\hat{\mathbf{X}}_\mathrm{Mask}$. The MSE loss $\mathcal{L}$ is then computed.}
    \label{fig:method}
    \vspace{-2mm}

\end{figure*}

\vspace{-2mm}

\section{Method}
\xhdr{Task Description}
Given a dataset of scRNA sequences $\mathbf{X}$, where $\mathbf{X}_{ij}$ is the expression value of $j$-th gene in $i$-th cell, we want to find a function $\mathcal{F}$ to generate an imputed scRNA seq matrix $\widetilde{\mathbf{X}}$ where $\widetilde{\mathbf{X}}_{ij}$ is the imputed expression value of $j$-th gene in $i$-th cell: $\mathcal{F}: \mathbf{X} \rightarrow \widetilde{\mathbf{X}}$. 

Since the dropout is unknown, we follow standard practice~\cite{arisdakessian2019deepimpute} to mask portions of non-zero values and make them zeros to simulate the dropout. This step and other data processing step is described in detail in Section.~\ref{sec:setup}. We denote the preprocessed scRNA-seq $\hat{\mathbf{x}}$. 

\xhdr{Induced hierarchical cell similarity graph} 
To aggregate gene information from similar cells, we construct a cell graph where the nodes are cells and edges connect similar cell. For a high quality graph, we filter the noise such as experiment artifacts by applying Principle Component Analysis (PCA) and using the top 50 dimensions to retain the majority of the variations. Then, we compute the euclidean distance among each cell-pair i and j, which accounts for their similarity. The similarity matrix $\mathbf{S}$ is constructed where each position $\mathbf{S}_{ij}$ is the similarity between cells $i$ and $j$. $\mathbf{S}$ is then used to construct a K-nearest neighbor (KNN) graph where each cell is connected to its K most-similar cells based on $\mathbf{S}$. This KNN graph is our cell similarity graph $G$ with adjacency matrix $\mathbf{A}$. 

Note that $G$ preserves the cell similarity hierarchies. For target cell $i$ that we want to impute, the immediate neighbors are the top K similar cells. Since we know for each neighbor $j$ of cell $i$, $j$ has K neighbors ($i$'s second-hop neighbors) that are the most similar to $j$. By the transitivity of euclidean distance, the second-hop neighbors are also similar to our target cell, although less similar to the first-hop neighbors. Hence, $G$ preserves the hierarchical structure of cell similarity. We call cells in the first-hop neighbors \textit{first-order} similar cells and second-hop neighbors \textit{second-order} similar cells and so forth.

\xhdr{Graph convolutional encoding}
We use a two-layer graph convolutional neural network~\cite{kipf2016semi} to aggregate the hierarchical cell similarity information in a simple propagation equation. GNN could be considered as message passing across connected nodes where messages are node attributes (scRNA sequence in our case). For the first layer of GCN, for any node cell $i$ with the attribute $\hat{\mathbf{X}}_i$, it aggregates node attributes from the connected neighbors $j \in \mathcal{N}(i)$ with a learnable matrix $\mathbf{W}_\mathbf{g}^1$ and a non-linear function $\sigma$ to obtain an updated cell representation $\mathbf{H}_i$ for target cell $i$:
\begin{equation*} 
    \mathbf{H}_i = \sigma \left(\frac{1}{\hat{\mathrm{K}}} \sum_{j \in \mathcal{N}(i)\cup i} \hat{\mathbf{X}}_j \mathbf{W}_\mathbf{g}^1 \right),
\end{equation*}
where $\hat{\mathrm{K}}$ is the KNN parameter K plus 1 for the node itself. It is also the degree of each cell in our similarity network. To interpret this, we see it takes the mean of the weighted first-order similar cells' scRNA-seq data. Intuitively, for dropout gene, this step imputes these gene expressions by taking the average of gene expressions from the most similar cells. For the second layer, the GCN applies the same propagation rule on top of the new $\mathbf{H}_i$ to obtain the imputed $\widetilde{\mathbf{X}}_i$:
\begin{equation*}
     \widetilde{\mathbf{X}}_i = \sigma \left(\frac{1}{\hat{\mathrm{K}}} \sum_{j \in \mathcal{N}(i)\cup i} \mathbf{H}_j \mathbf{W}_\mathbf{g}^2 \right).
\end{equation*}
The second propagation layer utilizes the second-order similar cells. It is clear when we combine the two propagation layers by expanding $\mathbf{H}_j$:
\begin{equation*} \label{eq:gcn}
    \widetilde{\mathbf{X}}_i = \sigma \left(\frac{1}{\hat{\mathrm{K}}} \sum_{j \in \mathcal{N}(i)\cup i} \left(\sigma \left(\frac{1}{\hat{\mathrm{K}}} \sum_{k \in \mathcal{N}(j)\cup j} \hat{\mathbf{X}}_k \mathbf{W}_\mathbf{g}^1 \right)\right) \mathbf{W}_\mathbf{g}^2 \right).
\end{equation*}
After the second layer, we see that GCN actually imputes the target gene expressions by aggregation first-order cells' expression by the factor $\frac{1}{\hat{\mathrm{K}}}$ and the second-order cells' expression by the factor $\frac{1}{\hat{\mathrm{K}}^2}$. Intuitively speaking, it aggregates similar cells where different level of similarity will have different aggregation weights. Less similar cells have less weights for the final imputation vector. This is an ideal property for dropout imputation that previous methods do not have. 

The hierarchical imputation equation can be vectorized for every cell in $G$ through matrix multiplication for fast paralleled computation~\cite{kipf2016semi}:
\begin{equation*} \label{eq:gcn_matrix}
    \widetilde{\mathbf{X}} = \sigma \left(\hat{\mathbf{A}}~\sigma \left( \hat{\mathbf{A}} \hat{\mathbf{X}} \mathbf{W}_\mathbf{g}^1 \right)\mathbf{W}_\mathbf{g}^2 \right)
\end{equation*}
where $\hat{\mathbf{A}} = \widetilde{\mathbf{D}}^{-\frac{1}{2}}\widetilde{\mathbf{A}}\widetilde{\mathbf{D}}^{-\frac{1}{2}}$, $\widetilde{\mathbf{D}}$ and $\widetilde{\mathbf{A}}$ are the renormalized degree and adjacency matrices of cell similarity graph $G$, defined as: $\widetilde{\mathbf{A}} = \mathbf{A} + \mathbf{I}$ and $\widetilde{\mathbf{D}}_{ii} = \sum_j \widetilde{\mathbf{A}}_{ij}$ ($\mathbf{I}$ is the identity matrix). We see this equation takes in the preprocessed data and outputs the imputed values.

\xhdr{Bottle-necked GCN weights}
We present a novel but simple modifications on GCN to leverage gene co-expression patterns. While the standard graph convolutional network enables aggregation over similar cells, to leverage gene co-expression pattern is not obvious. Previous successful methods~\cite{arisdakessian2019deepimpute,talwar2018autoimpute,van2018recovering} use auto-encoder to first map the scRNA-seq into a low dimensional latent space and then recover it to the original space. We borrow this idea of bottleneck into the GCN framework. A brute-force adaptation would be to include a separate auto-encoder and concatenate the latent embedding from both auto-encoder and GCNs. However, this requires additional computation memory and speed. Instead, we directly inject this bottleneck into the GCN framework by setting the weight matrix $\mathbf{W}_\mathbf{g}^1$ to have dimension $\mathbb{R}^{M \times D}$ and weight matrix $\mathbf{W}_\mathbf{g}^2$ to have dimension  $\mathbb{R}^{D \times M}$, where $D << M$ ($M$ is the number of genes in the input). This way, we inject an auto-encoder into the GCN without any additional computation time and it has the same effect of auto-encoder since the weight matrix is applied on every cells. 

\xhdr{Loss and inference} 
Given the input $\hat{\mathbf{X}}$ and the output $\widetilde{\mathbf{X}}$, we first retrieve the corresponding masked values $\hat{\mathbf{X}}_\mathrm{Mask}$ and $\widetilde{\mathbf{X}}_\mathrm{Mask}$. Then, we compute the reconstruction loss: $ \mathcal{L} = \Vert \hat{\mathbf{X}}_\mathrm{Mask} - \widetilde{\mathbf{X}}_\mathrm{Mask} \Vert_2 $. This loss measures the accuracy of dropout imputation on the masked true gene expressions. During inference, we fill in the masking values and freeze the model weights. Then, we put the predicted values for the dropout position while leave the non-zero values intact. The imputed scRNA-seq is then obtained and ready for downstream biomedical analysis.
\vspace{-2mm}
\section{Experiment}
\begin{table}[t]
    \centering   
    \caption{Dataset Statistics.}

    \begin{tabular}{l|ccc}
    \toprule
    Dataset & \# Cells & \#Genes & \% Mask  \\ \hline
    Zeisel & 3,005 & 19,972 & 18.79\\
    Neuron1K & 1,301 & 31,053 & 10.45 \\
    Jurkat  & 3,258 & 32,738 & 9.76 \\
    \bottomrule
    \end{tabular}
    \label{tab:2}
\end{table}

\subsection{Setup} \label{sec:setup}
\begin{figure}[t]
    \centering
    \includegraphics[width = 0.49\textwidth]{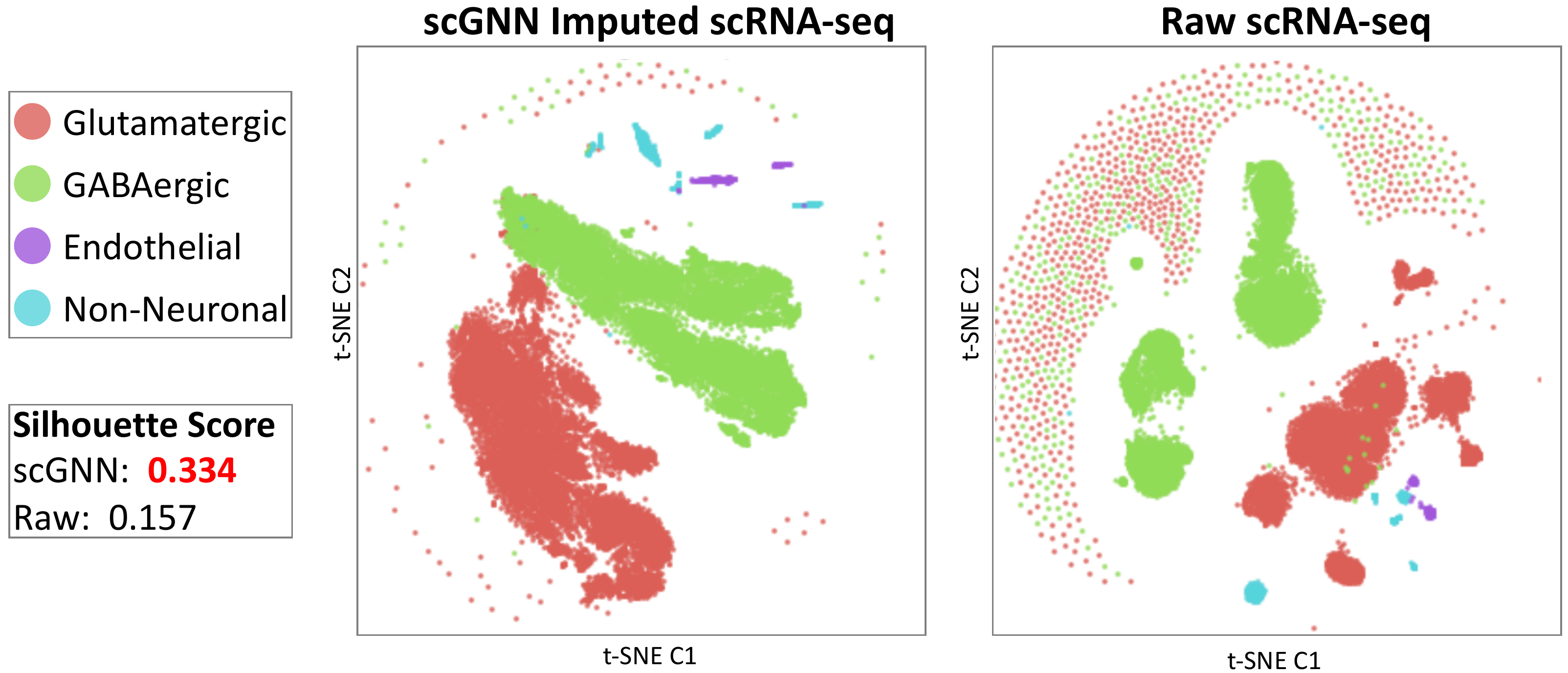}
        \vspace{-3mm}
    \caption{\textbf{\name improves downstream biomedical analysis.} It obtains better clustering result than the raw scRNA-seq data, measured by more than 100\% increase on clustering silhouette score. }
    \label{fig:3}
    \vspace{-3mm}
\end{figure}
\xhdr{Datasets and Data Processing}
We use three datasets. The first one is Zeisel dataset~\cite{zeisel2015cell}, which is scRNA-seq from mouse cortex and hippocampus and has 3,005 cells and 19,972 genes. The second is Neuron1K. It derives from the brain cells from E18 mouses~(10X Genomics). It has 1,301 cells and 31,053 genes. The third one is Jurkat~(10X Genomics), which is extracted from human blood cell lines and have 3,258 cells and 32,738 genes.

There are three steps for the standard data processing~\cite{talwar2018autoimpute,arisdakessian2019deepimpute}. First we select important genes as many genes are not useful. We filter genes that have mean over variance ratio (MVR) less than 0.5. Second, we conduct median normalization of transcript abundance by forcing each cell has the same transcript counts as the median across cells. This way we eliminate cell size as the potential inaccurate imputation. In the last, we calculate the dropout percentage in the original scRNA-seq dataset and randomly mask the same percentage on the processed dataset.

\xhdr{Evaluation Metrics}
We use two metrics for evaluating dropout imputation accuracy. The first one is Mean Squared Error (MSE), which equals to the loss function. The second one is the Pearson Correlation ($R^2$), which measures the correlation between the predicted and true transcript counts.

\xhdr{Baselines}
To compare \name's performance, we evaluate dropout imputation performance for 1. \textbf{MAGIC}~\cite{van2018recovering} constructs Markov affinity matrix to model cell similarity and 2. \textbf{DCA}~\cite{eraslan2019single} uses denoising auto-encoders, taking account into the count distribution, overdispersion and sparsity of the data by constructing a negative binomial loss. 
\vspace{-2mm}
\subsection{Preliminary Result}

\xhdr{\name\footnote{Source code available at \url{https://github.com/kexinhuang12345/scGNN}.} is a robust, competitive dropout imputation algorithm}
The prediction result measured by R-squared and MSE is reported in Figure.~(\ref{tab:result}). We see that in Zeisel and Neuron1K, \name has the highest R-squared and lowest MSE. In Jurkat, \name's performance is worse than DCA and MAGIC in R-Squared and better than DCA in MSE. This suggests \name has very good predictive performances, and in often times, better predictions. We see both DCA and MAGIC has higher variance in all the three datasest, especially MAGIC, it could not infer high-quality dropouts in the Zeisel dataset, suggested by the mediocre R-squared result. In the Neuron1K dataset, we found DCA and MAGIC perform significantly bad in one random data split fold while \name obtains consistently good result, suggesting \name is robust.

\xhdr{\name improves downstream function clustering analysis}
The purpose of doing dropout imputation is to enable more accurate downstream scRNA-seq analysis. In this experiment, we compare the cell types t-SNE visualization, before and after \name imputation. We report the result in Figure.~(\ref{fig:3}). We see that before \name imputation, the Glutamatergic and GABAergic, two clusters, are separated into several small parts. \name is able to identify the main variance and cluster them together. To quantify the improvement, we use the silhouette score, which measures the how similar a cell is to its own cluster compared to other clusters. We compute the scores, raw data achieves 0.157, and \name achieves 0.334, a nearly 100\% relative increase over raw data. This suggests the benefit of using \name and its usefulness in improving downstream biomedical analysis. 

\begin{table}[t]
    \centering
    \caption{Five fold average and standard deviation of MSE and R2.}
    \adjustbox{max width= 0.49 \textwidth}{
    \begin{tabular}{l|cc|cc|cc}
    \toprule
    Dataset & \multicolumn{2}{c|}{Zeisel} & \multicolumn{2}{c|}{Neuron1K} & \multicolumn{2}{c}{Jurkat} \\ \midrule
    Metric & MSE & R2 & MSE & R2 & MSE & R2 \\ \midrule
    DCA  & 165.7 & 0.863 & 411.9 & 0.931 & 40.3 & 0.960  \\ 
    MAGIC & 348.6 & 0.536 & 441.8 & 0.825 & 22.1 & 0.962 \\
    \name & 98.6 & 0.871 & 113.0 & 0.943 & 32.7 & 0.933 \\
    \bottomrule
    \end{tabular}}
    \label{tab:result}
\end{table}

\section{Conclusion}
In this work, we present a novel, simple, and efficient computational method that unifies cell similarity and gene co-expression patterns and also leverages the hierarchical nature of cell similarity. \name achieves great dropout imputation predictive performance and improves downstream analysis.

\bibliographystyle{icml2019}
\bibliography{ref} 

\begin{thebibliography}{15}
\providecommand{\natexlab}[1]{#1}
\providecommand{\url}[1]{\texttt{#1}}
\expandafter\ifx\csname urlstyle\endcsname\relax
  \providecommand{\doi}[1]{doi: #1}\else
  \providecommand{\doi}{doi: \begingroup \urlstyle{rm}\Url}\fi

\bibitem[Arisdakessian et~al.(2019)Arisdakessian, Poirion, Yunits, Zhu, and
  Garmire]{arisdakessian2019deepimpute}
Arisdakessian, C., Poirion, O., Yunits, B., Zhu, X., and Garmire, L.~X.
\newblock Deepimpute: an accurate, fast, and scalable deep neural network
  method to impute single-cell rna-seq data.
\newblock \emph{Genome biology}, 20\penalty0 (1):\penalty0 1--14, 2019.

\bibitem[Eraslan et~al.(2019)Eraslan, Simon, Mircea, Mueller, and
  Theis]{eraslan2019single}
Eraslan, G., Simon, L.~M., Mircea, M., Mueller, N.~S., and Theis, F.~J.
\newblock Single-cell rna-seq denoising using a deep count autoencoder.
\newblock \emph{Nature communications}, 10\penalty0 (1):\penalty0 1--14, 2019.

\bibitem[Gladka et~al.(2018)Gladka, Molenaar, De~Ruiter, Van Der~Elst, Tsui,
  Versteeg, Lacraz, Huibers, Van~Oudenaarden, and Van~Rooij]{gladka2018single}
Gladka, M.~M., Molenaar, B., De~Ruiter, H., Van Der~Elst, S., Tsui, H.,
  Versteeg, D., Lacraz, G.~P., Huibers, M.~M., Van~Oudenaarden, A., and
  Van~Rooij, E.
\newblock Single-cell sequencing of the healthy and diseased heart reveals
  cytoskeleton-associated protein 4 as a new modulator of fibroblasts
  activation.
\newblock \emph{Circulation}, 138\penalty0 (2):\penalty0 166--180, 2018.

\bibitem[Gong et~al.(2018)Gong, Kwak, Pota, Koyano-Nakagawa, and
  Garry]{gong2018drimpute}
Gong, W., Kwak, I.-Y., Pota, P., Koyano-Nakagawa, N., and Garry, D.~J.
\newblock Drimpute: imputing dropout events in single cell rna sequencing data.
\newblock \emph{BMC bioinformatics}, 19\penalty0 (1):\penalty0 220, 2018.

\bibitem[Hicks et~al.(2018)Hicks, Townes, Teng, and Irizarry]{hicks2018missing}
Hicks, S.~C., Townes, F.~W., Teng, M., and Irizarry, R.~A.
\newblock Missing data and technical variability in single-cell rna-sequencing
  experiments.
\newblock \emph{Biostatistics}, 19\penalty0 (4):\penalty0 562--578, 2018.

\bibitem[Huang et~al.(2018)Huang, Wang, Torre, Dueck, Shaffer, Bonasio, Murray,
  Raj, Li, and Zhang]{huang2018saver}
Huang, M., Wang, J., Torre, E., Dueck, H., Shaffer, S., Bonasio, R., Murray,
  J.~I., Raj, A., Li, M., and Zhang, N.~R.
\newblock Saver: gene expression recovery for single-cell rna sequencing.
\newblock \emph{Nature methods}, 15\penalty0 (7):\penalty0 539--542, 2018.

\bibitem[Jia et~al.(2017)Jia, Hu, Kelly, Kim, Li, and Zhang]{jia2017accounting}
Jia, C., Hu, Y., Kelly, D., Kim, J., Li, M., and Zhang, N.~R.
\newblock Accounting for technical noise in differential expression analysis of
  single-cell rna sequencing data.
\newblock \emph{Nucleic acids research}, 45\penalty0 (19):\penalty0
  10978--10988, 2017.

\bibitem[Keren-Shaul et~al.(2017)Keren-Shaul, Spinrad, Weiner,
  Matcovitch-Natan, Dvir-Szternfeld, Ulland, David, Baruch, Lara-Astaiso, Toth,
  et~al.]{keren2017unique}
Keren-Shaul, H., Spinrad, A., Weiner, A., Matcovitch-Natan, O.,
  Dvir-Szternfeld, R., Ulland, T.~K., David, E., Baruch, K., Lara-Astaiso, D.,
  Toth, B., et~al.
\newblock A unique microglia type associated with restricting development of
  alzheimer’s disease.
\newblock \emph{Cell}, 169\penalty0 (7):\penalty0 1276--1290, 2017.

\bibitem[Kharchenko et~al.(2014)Kharchenko, Silberstein, and
  Scadden]{kharchenko2014bayesian}
Kharchenko, P.~V., Silberstein, L., and Scadden, D.~T.
\newblock Bayesian approach to single-cell differential expression analysis.
\newblock \emph{Nature methods}, 11\penalty0 (7):\penalty0 740, 2014.

\bibitem[Kipf \& Welling(2016)Kipf and Welling]{kipf2016semi}
Kipf, T.~N. and Welling, M.
\newblock Semi-supervised classification with graph convolutional networks.
\newblock \emph{arXiv preprint arXiv:1609.02907}, 2016.

\bibitem[Stephenson et~al.(2018)Stephenson, Donlin, Butler, Rozo, Bracken,
  Rashidfarrokhi, Goodman, Ivashkiv, Bykerk, Orange,
  et~al.]{stephenson2018single}
Stephenson, W., Donlin, L.~T., Butler, A., Rozo, C., Bracken, B.,
  Rashidfarrokhi, A., Goodman, S.~M., Ivashkiv, L.~B., Bykerk, V.~P., Orange,
  D.~E., et~al.
\newblock Single-cell rna-seq of rheumatoid arthritis synovial tissue using
  low-cost microfluidic instrumentation.
\newblock \emph{Nature communications}, 9\penalty0 (1):\penalty0 1--10, 2018.

\bibitem[Talwar et~al.(2018)Talwar, Mongia, Sengupta, and
  Majumdar]{talwar2018autoimpute}
Talwar, D., Mongia, A., Sengupta, D., and Majumdar, A.
\newblock Autoimpute: Autoencoder based imputation of single-cell rna-seq data.
\newblock \emph{Scientific reports}, 8\penalty0 (1):\penalty0 1--11, 2018.

\bibitem[Usoskin et~al.(2015)Usoskin, Furlan, Islam, Abdo, L{\"o}nnerberg, Lou,
  Hjerling-Leffler, Haeggstr{\"o}m, Kharchenko, Kharchenko,
  et~al.]{usoskin2015unbiased}
Usoskin, D., Furlan, A., Islam, S., Abdo, H., L{\"o}nnerberg, P., Lou, D.,
  Hjerling-Leffler, J., Haeggstr{\"o}m, J., Kharchenko, O., Kharchenko, P.~V.,
  et~al.
\newblock Unbiased classification of sensory neuron types by large-scale
  single-cell rna sequencing.
\newblock \emph{Nature neuroscience}, 18\penalty0 (1):\penalty0 145, 2015.

\bibitem[Van~Dijk et~al.(2018)Van~Dijk, Sharma, Nainys, Yim, Kathail, Carr,
  Burdziak, Moon, Chaffer, Pattabiraman, et~al.]{van2018recovering}
Van~Dijk, D., Sharma, R., Nainys, J., Yim, K., Kathail, P., Carr, A.~J.,
  Burdziak, C., Moon, K.~R., Chaffer, C.~L., Pattabiraman, D., et~al.
\newblock Recovering gene interactions from single-cell data using data
  diffusion.
\newblock \emph{Cell}, 174\penalty0 (3):\penalty0 716--729, 2018.

\bibitem[Zeisel et~al.(2015)Zeisel, Mu{\~n}oz-Manchado, Codeluppi,
  L{\"o}nnerberg, La~Manno, Jur{\'e}us, Marques, Munguba, He, Betsholtz,
  et~al.]{zeisel2015cell}
Zeisel, A., Mu{\~n}oz-Manchado, A.~B., Codeluppi, S., L{\"o}nnerberg, P.,
  La~Manno, G., Jur{\'e}us, A., Marques, S., Munguba, H., He, L., Betsholtz,
  C., et~al.
\newblock Cell types in the mouse cortex and hippocampus revealed by
  single-cell rna-seq.
\newblock \emph{Science}, 347\penalty0 (6226):\penalty0 1138--1142, 2015.

\end{thebibliography}

\end{document}